\documentclass[twocolumn,secnumarabic,superscriptaddress,nobibnotes,aps,prd,showpacs]{revtex4}

\usepackage{color}
\usepackage{graphicx}
\usepackage{bm}
\usepackage{amsmath}
\usepackage{epsfig}
\usepackage{amsfonts}
\usepackage{amssymb}
\usepackage{cancel}
\usepackage{dsfont}
\usepackage{multirow}

\begin{document}
	
	\title{How the axial anomaly controls flavor mixing among mesons}
	
	\author{Francesco Giacosa}
	\email{fgiacosa@ujk.edu.pl}
	\affiliation{Institute~of~Physics, Jan~Kochanowski~University, ul.~Swietokrzyska 15, 25-406 Kielce}
	\affiliation{Institute~for~Theoretical~Physics, Johann~Wolfgang~Goethe~University, Max-von-Laue-Str.~1, 60438 Frankfurt~am~Main}
	
	\author{Adrian Koenigstein}
	\email{koenigstein@th.physik.uni-frankfurt.de}
	\affiliation{Institute~for~Theoretical~Physics, Johann~Wolfgang~Goethe~University, Max-von-Laue-Str.~1, 60438 Frankfurt~am~Main}
	
	\author{Robert D. Pisarski}
	\email{pisarski@bnl.gov}
	\affiliation{Department~of~Physics, Brookhaven~National~Laboratory, Upton, NY 11973}
	
	\date{\today}

\begin{abstract}
	
	It is well known that because of the axial anomaly in QCD, mesons with $J^P = 0^-$ are close to $SU(3)_{\mathrm{V}}$ eigenstates: the $eta^\prime(958)$ meson is largely a singlet, and the $\eta$ meson an octet. In contrast, states with $J^P = 1^-$ are flavor diagonal: \textit{e.g.}, the $\phi(1020)$ is almost pure $\bar{s} s$. Using effective Lagrangians, we show how this generalizes to states with higher spin, assuming that they can be classified according to the unbroken chiral symmetry of $G_{\mathrm{fl}} = SU(3)_{\mathrm{L}} \times SU(3)_{\mathrm{R}}$. We construct effective Lagrangians from terms invariant under $G_{\mathrm{fl}}$, and introduce the concept of \textit{hetero-} and \textit{homo}chiral multiplets. Because of the axial anomaly, only terms invariant under the $Z(3)_{\mathrm{A}}$ subgroup of the axial $U(1)_{\mathrm{A}}$ enter. For heterochiral multiplets, which begin with that including the $\eta$ and $\eta^\prime(958)$, there are $Z(3)_{\mathrm{A}}$ invariant terms with low mass dimension which cause states to mix according to $SU(3)_{\mathrm{V}}$ flavor. For homochiral multiplets, which begin with that including the $\phi(1020)$, there are no $Z(3)_{\mathrm{A}}$ invariant terms with low mass dimension, and so states are diagonal in flavor. In this way we predict the flavor mixing for the heterochiral multiplet with spin one, as well as for hetero- and homochiral multiplets with spin two and spin three.
	
\end{abstract}

	
	\maketitle
	



	\textit{Introduction} -- The theory of quarks and gluons, Quantum chromodynamics (QCD), has two symmetries. There is an exact local symmetry of $SU(3)_\mathrm{C}$ color, which is responsible for the confinement of quarks and gluons into hadrons. For the light quark flavors of up, down, and strange quarks, ($u$, $d$, and $s$) there is also an approximate chiral symmetry \cite{Thomas:2001kw}. This is spontaneously broken in vacuum to a residual symmetry of $SU(3)_\mathrm{V}$ flavor. The up and down quarks are much lighter than hadronic scales, with the breaking of $SU(3)_\mathrm{V}$ flavor dominated by the effects of the strange quark, whose mass is comparable to hadronic scales. Chiral symmetry is responsible for the most evident feature of hadrons, that when it is spontaneously broken in vacuum, there is an octet of light pseudo-Goldstone bosons with spin parity $J^P = 0^-$, which are pions, kaons, and the $\eta$ meson. Because of the mass of the strange quark, kaons, which include strange quarks, are heavier than pions, which do not.
	
	This leaves a puzzle: why isn't there a ninth light state, a $SU(3)_\mathrm{V}$ singlet? Instead, the corresponding state, the $\eta^\prime(958)$ meson, is unexpectedly heavy. Indeed, why should the pseudo-Goldstone bosons form states in representations of $SU(3)_\mathrm{V}$ flavor? Since the strange quark is so much heavier than the up and down quarks, why aren't the $\pi^0$ and $\eta$ composed just of up and down quarks, and the $\eta^\prime(958)$ purely strange \cite{Gross:1979ur}? The solution to these puzzles involves a global, axial symmetry of $U(1)_\mathrm{A}$, which while valid classically, is reduced by quantum effects to a discrete subgroup of $Z(3)_\mathrm{A}$ \cite{tHooft:1986ooh}. This reduction in symmetry both pushes the mass of the \textcolor{blue}{$\eta^\prime(958)$} up, and forces it to be close to a singlet under $SU(3)_\mathrm{V}$ flavor \cite{Gross:1979ur}. (In fact, the $\eta_N = (\bar{u}u + \bar{d}d)/\sqrt{2}$ and $\eta_S = \bar{s}s$ do mix because of the strange quark mass.) This can be demonstrated in effective models, either nonlinear or linear \cite{Thomas:2001kw,tHooft:1986ooh,Gross:1979ur,Feldmann:1998vh}.
	
	What about other hadronic states? The lightest vector multiplet has $J^P = 1^-$. This includes the $\rho(770)$ and $\omega(782)$ mesons, which are composed almost exclusively of up and down quarks, and the $\phi(1020)$ meson, which has mainly strange quarks. This difference can also be understood in effective models, such as through the Wess-Zumino-Witten Lagrangian, where effects due to the axial anomaly do not appear in mass terms, but only those of relatively high mass dimension.
	
	This leads to the question which we address in this work: how does this striking difference in flavor mixing show up in hadronic multiplets with \textit{higher spin}? We consider the case of the multiplet with $J^P= 1^+$, and multiplets with spin two and three. At the outset, we stress that our basic assumption is that we can classify the transformation properties of these multiplets according to the \textit{unbroken} chiral symmetry. This is not self evident for such heavy states, which experimentally have masses between $1$ and $2$ GeV. Nevertheless, this assumption allows us to make numerous predictions for their masses, flavor mixing, and decay modes. While most of the states which we discuss are not well measured experimentally, we hope that our comments might contribute to their further study.
	
	In this work, as it is convenient, we chose an assignment for some of the mesons below $2$ GeV. We assume that the resonances under consideration are (at least predominantly) $\bar{q}q$ states. This is not trivial, since in principle each resonance is not a simple quark-anitquark object but a superposition of various components, which include meson-meson bound states, tetraquarks, gluons, etc.. Moreover, different approaches give rise to different interpretations of many resonances, e.g.\ \cite{Amsler:1995tu,Klempt:2007cp,Pelaez:2015qba}. For definiteness, here we use the assignment presented in the `Quark Model' review of the Particle Data Group \cite{pdg}. While the nature of some resonances is disputed and certain parts of this assignment may change in the future, it gives us a valuable basis to start with. Moreover, our theoretical considerations  still apply, even if some of the specific assignments may change.
	

	\textit{Chiral symmetry} -- Suppressing color (indices), we denote quark fields as $q^i=(u,d,s)$. Left- and right-handed quarks fields are defined as
		\begin{align}
			&q_{\mathrm{L},\mathrm{R}} = \mathcal{P}_{\mathrm{L},\mathrm{R}}\, q\,, &&\bar{q}_{\mathrm{L},\mathrm{R}} = \bar{q}\, \mathcal{P}_{\mathrm{R},\mathrm{L}}\,, \label{defquark}
		\end{align}
                where $\mathcal{P}_{\mathrm{L},\mathrm{R}} = (\mathds{1} \mp \gamma^5)/2$ are orthogonal projectors, with $\mathcal{P}_{\mathrm{L},\mathrm{R}}^2 = \mathcal{P}_{\mathrm{L},\mathrm{R}}$ and $\mathcal{P}_\mathrm{L} \mathcal{P}_\mathrm{R} = 0$, taking $(\gamma ^{5})^2 = \mathds{1}$. Notice that because $\bar{q} = q^\dagger \gamma^0$ in Minkowski spacetime, in Eq.\ (\ref{defquark}) the antiquark has a projector opposite to that of the quark. Under the global chiral symmetry of
                $G_\mathrm{fl} \times U(1)_\mathrm{A} = SU(3)_\mathrm{L} \times SU(3)_\mathrm{R} \times U(1)_\mathrm{A}$,
		\begin{align}
			q_{\mathrm{L},\mathrm{R}} \longrightarrow \mathrm{e}^{\mp \mathrm{i} \alpha/2}\, U_{\mathrm{L},\mathrm{R}}\, q_{\mathrm{L},\mathrm{R}}\,,
		\end{align}
	where $U_{\mathrm{L},\mathrm{R}}$ are rotations in $SU(3)_{\mathrm{L},\mathrm{R}}$ and $\alpha$ is the parameter for an axial rotation in $U(1)_\mathrm{A}$.
	
	The flavor symmetry $G_\mathrm{fl}$ is exact in the limit that all quark masses vanish. It is spontaneously broken in the QCD vacuum to the diagonal subgroup of $SU(3)_{\mathrm{V}}$, under which $U_{\mathrm{L}} = U_{\mathrm{R}} = U_{\mathrm{V}}$. This generates an octet of massless Goldstone bosons.
	
	The flavor symmetry is only approximate once current quark masses are included. In QCD the masses of the up, down, and strange quarks are $m_u \sim 5$, $m_d \sim 10$, and $m_s \sim 100$~MeV, respectively. Symmetry breaking generates pseudo-Goldstone bosons, with the mass squared of the pion $m_\pi^2 \propto ( m_u + m_d )$, and that of the kaon, $m_K^2 \propto ( m_u + m_s )$.
	
	The axial symmetry is special. While the currents for $G_{\mathrm{fl}}$ are conserved in the limit of vanishing quark masses, that for the axial $U(1)_{\mathrm{A}}$ is proportional to the topological charge density \cite{tHooft:1986ooh,Adler:1969gk,Bell:1969ts},
		\begin{align}
			\partial^\mu ( \bar{q}^i \gamma_\mu \gamma_5 q^i ) = \frac{3 g^2}{16\pi^2}\, \varepsilon^{\mu\nu\rho\sigma} \mathrm{tr} ( G_{\mu\nu} G_{\rho\sigma})\,.
		\end{align}
	This reduces the $U(1)_\mathrm{A}$ symmetry to one of $Z(3)_\mathrm{A}$. This residual symmetry can be understood from the vertex generated by the zero modes of a single instanton, which for three flavors couples $u_\mathrm{L}$, $d_\mathrm{L}$, and $s_\mathrm{L}$ with $\bar{u}_\mathrm{R}$, $\bar{d}_\mathrm{R}$, and $\bar{s}_\mathrm{R}$. Under $q_{\mathrm{L},\mathrm{R}} \rightarrow \mathrm{e}^{\mp \mathrm{i} \pi/3}\, q_{\mathrm{L},\mathrm{R}}$ this vertex changes as $\mathrm{e}^{2 \pi \mathrm{i}}$, and so is $Z(3)_\mathrm{A}$ invariant. The same applies to all multi-instanton interactions.
	
	
	\textit{Heterochiral scalars} -- We begin by reviewing the effect of the axial anomaly on scalar fields. Because of confinement, we form fields which are color singlets (and so always implicitly sum over color indices), but transform non-trivially under flavor transformations. The simplest is to form a scalar by pairing a quark and an anti-quark. Since the chiral projectors are orthogonal, pairing a quark and anti-quark with the same chirality automatically vanishes, \textit{e.g.}, $\bar{q}_{\mathrm{L}}\, q_{\mathrm{L}} = \bar{q}\, \mathcal{P}_{\mathrm{R}} \mathcal{P}_{\mathrm{L}}\, q = 0$. Instead, we must take
		\begin{align}
			\Phi^{ij} \equiv \bar{q}^j \mathcal{P}_\mathrm{L} q^i = \bar{q}_\mathrm{R}^j q_\mathrm{L}^i\,,
		\end{align}
	with $\Phi^\dagger = \bar{q}_\mathrm{L} q_\mathrm{R}$. This field transforms under $G_\mathrm{fl} \times U(1)_\mathrm{A}$ as
		\begin{align}
			\Phi \longrightarrow \mathrm{e}^{-\mathrm{i}\alpha}\, U_\mathrm{L}\, \Phi\, U_\mathrm{R}^\dagger\,. \label{chiral_transf_phi}
		\end{align}
	Since we must pair fields with opposite chirality, we term $\Phi $ as a \textit{heterochiral} field. Consequently, the transformation of $\Phi$ under the flavor group involves both $U_\mathrm{L}$ and $U_\mathrm{R}$.
	
	The components of $\Phi$ with $J^P = 0^-$, given by $P = (\Phi - \Phi^\dagger)/2\mathrm{i}$, are the $\pi$, $K$, $\eta$, and $\eta^\prime(958)$; those with $J^P = 0^+$, denoted as $S = (\Phi + \Phi^\dagger)/2$, are assigned to be the $a_0(1450)$, $K_0^*(1430)$, $f_0(1370)$, and $f_0(1710)$. In the quark model their quantum numbers are ${}^{2S+1}L_J = {}^{1}S_0$ and ${}^{3}P_0$, respectively.
	
	Here, we assume that the $0^+$ states below $1$ GeV are predominantly four-quark states \cite{Amsler:1995tu,Klempt:2007cp,Pelaez:2015qba,Ochs:2013gi,Janowski:2014ppa,Gui:2012gx,Brunner:2015oqa}, but this is really secondary to our analysis. However, the non-standard nature of the light scalar states is confirmed by various works, e.g.\ \cite{Amsler:1995tu,Klempt:2007cp,Pelaez:2015qba,Ochs:2013gi,Janowski:2014ppa} and Refs.\ therein, where they can be obtained as dynamically generated `molecular-like' states. In agreement with this view, the large-$N_\mathrm{C}$ behavior of the corresponding pole confirms this feature \cite{Pelaez:2003dy}. Yet, mixing is possible, hence a certain $\bar{q}q$ amount into light scalars is present, but, as argued above, we regard it as subleading. While in this work we concentrate on the states which are predominantly $\bar{q}q,$ the axial anomaly can play a role also for four-quark states \cite{Pisarski:2016ixt}: we leave a detailed study of this issue for the future. Similarly, the above listed scalar states above $1$ GeV are not simple $\bar{q}q$ resonances; mixing with four-quark components and with a scalar glueball are expected \cite{Geng:2008gx,Ochs:2013gi,Janowski:2014ppa,Giacosa:2009bj,Fariborz:2005gm,Napsuciale:2004au}. Due to all these complications, we do not attempt to study the anomaly-driven mixing of the scalar-isoscalar $\bar{q}q$ components.
	
	When spontaneous symmetry breaking occurs, we can take the expectation value of $\Phi$ to be approximately diagonal in flavor, $\langle \Phi \rangle = \sqrt{3/2}\, \phi_0 \, \mathds{1}$, where $\phi_0 = f_\pi \sim 93$~MeV, which is a consequence of flavor symmetry.
	
	In constructing an effective model of QCD at low energies, one includes all terms which are symmetric under $G_\mathrm{fl}$ \cite{Gasiorowicz:1969kn,Rosenzweig:1979ay,Fariborz:2005gm,Pisarski:1994yp,Pisarski:2016ixt,Parganlija:2012fy,Parganlija:2010fz}. The simplest is to take traces of powers of $\Phi^\dagger \Phi$, such as $\mathrm{tr}(\Phi^\dagger \Phi )$, $\mathrm{tr}(\Phi^\dagger \Phi)^2$, \textit{etc.}. These are all invariant under $U(1)_\mathrm{A}$. Terms which are only invariant under $Z(3)_\mathrm{A}$ begin with $\det(\Phi)$. This is generated by the zero modes of an instanton, and so transforms as $\det(\Phi) \rightarrow \mathrm{e}^{- 3 \mathrm{i} \alpha} \det(\Phi)$ under $U(1)_\mathrm{A}$ \cite{tHooft:1986ooh}. The polynomial terms with the lowest mass dimension which are $Z(3)_\mathrm{A}$ invariant, Hermitian, and parity invariant are
		\begin{align}
			\mathcal{L}_\Phi^{\text{anomaly}} = & - a_\mathrm{A}^{(1)} [ \mathrm{det}(\Phi) + \mathrm{c.c.} ] - a_\mathrm{A}^{(2)} [\mathrm{det}(\Phi) + \mathrm{c.c.}]^2 \nonumber
			\\
			& - a_\mathrm{A}^{(3)} [\mathrm{det}(\Phi) - \mathrm{c.c.}]^2 - \ldots \label{anomalypseudoscalar}
		\end{align}
	The first term is cubic in the fields, and when $\phi_0 \neq 0$, drives the mass for the singlet pseudo-scalar up, and that for the singlet scalar down \cite{tHooft:1986ooh,Pisarski:1994yp,Parganlija:2010fz,Pisarski:2016ixt,Kovacs:2016juc}. The third term affects only isoscalar-pseudoscalar states, contributing to their mass squared as
		\begin{align}
			- \alpha_\mathrm{A} \eta_0^2 & = - \frac{\alpha_\mathrm{A}}{3} (\sqrt{2} \eta_N + \eta_S)^2 \nonumber
			\\
			& = -\frac{\alpha_\mathrm{A}}{3} (2 \eta_N^2 + \eta_S^2 + 2 \sqrt{2} \eta_N \eta_S)\,,
		\end{align}
	where $\alpha_\mathrm{A} \sim a_\mathrm{A}^{(3)} \phi _0^4$ \cite{Parganlija:2012fy,Rosenzweig:1979ay,Fariborz:2005gm}.
It is in agreement with the Witten-Veneziano mass formula \cite{Veneziano:1979ec,Witten:1980sp}, and arises naturally when integrating out a pseudoscalar glueball \cite{Eshraim:2012jv}. The term $\sim a_\mathrm{A}^{(3)}$ contributes to the mixing angle between the $\eta_N$ and the $\eta_S$ as $2 \theta_P = - \arctan( 2 \sqrt{2} \alpha_\mathrm{A} / ( m_K^2 - m_\pi^2 - \alpha_\mathrm{A} )$. For realistic values of $\alpha_\mathrm{A}$ \cite{Bass:2005hn}, this explains the negative value of $\theta_P \sim - 42^\circ$ observed experimentally \cite{AmelinoCamelia:2010me}. The second term in Eq.\ (\ref{anomalypseudoscalar}) complicates the analysis. See also \cite{Escribano:2007cd,AmelinoCamelia:2010me,Feldmann:1998vh,Borasoy:2005du} for further phenomenological studies on the $\eta$ and $eta^\prime(958)$.
	
	
	\textit{Homochiral vectors} -- Vector mesons are formed by pairing a quark and an anti-quark with a Dirac matrix $\gamma_\mu$. Since $\gamma^5$ anti-commutes with $\gamma_\mu$, we must pair a quark and anti-quark with the \textit{same} chirality
		\begin{align}
			&L_\mu^{ij} \equiv \bar{q}_\mathrm{L}^j \gamma_\mu q_\mathrm{L}^i\,, &&R_\mu^{ij} \equiv \bar{q}_\mathrm{R}^j \gamma_\mu q_\mathrm{R}^i\,.
		\end{align}
	The left- and right-handed vector fields transform under the flavor group $G_\mathrm{fl} \times U(1)_\mathrm{A}$ as
		\begin{align}
			&L_\mu \longrightarrow U_\mathrm{L}\, L_\mu\, U_\mathrm{L}^\dagger\,, &&R_\mu \longrightarrow U_\mathrm{R}\, R_\mu\, U_\mathrm{R}^\dagger\,. \label{homo_vector_transf}
		\end{align}
	We term these fields as \textit{homochiral}, with $L_\mu^{ij}$ transforming as a two-index tensor under $SU(3)_\mathrm{L}$, and similarly for $R_\mu^{ij}$ under $SU(3)_\mathrm{R}$.
	
	The homochiral $L_\mu$ and $R_\mu$ only couple to the heterochiral $\Phi$ and $\Phi^\dagger$ through terms which involve derivatives of $\Phi$. The first such term is
		\begin{align}
			\mathrm{tr} [L_\mu \Phi (\partial^\mu \Phi^\dagger) + R_\mu \Phi^\dagger (\partial^\mu \Phi)]\,. \label{vector_mixing}
		\end{align}
	When $\phi_0 \neq 0$ this term causes the homochiral vector fields to mix with derivatives of $\Phi$.
	
	The vector fields themselves are automatically invariant under the axial $U(1)_\mathrm{A}$ symmetry, and so it is not possible to write a term which involves the axial anomaly using only these fields. Such effects are included by the coupling of the vector fields to the $\Phi$'s \cite{Gomm:1984at} , such as through Wess-Zumino-Witten type terms,
		\begin{align}
			\varepsilon^{\mu\nu\alpha\beta} \mathrm{tr}[ &L_\mu \Phi (\partial_\nu \Phi^\dagger) \Phi (\partial_\alpha \Phi^\dagger) \Phi (\partial_\beta \Phi^\dagger) \nonumber
			\\
			&+ R_\mu \Phi^\dagger (\partial_\nu \Phi) \Phi^\dagger (\partial_\alpha \Phi) \Phi^\dagger (\partial_\beta \Phi) ]\,.
		\end{align}
	Experimentally the homochiral vector multiplet is well known. The vectors with negative parity, $J^P = 1^-$, are given by $V_\mu = (L_\mu + R_\mu)/2$: these are the $\rho(770)$, $K^*(892)$, $\omega(782)$, and $\phi(1020)$. Axial-vectors with positive parity, $J^P = 1^+$, are $A_\mu = (L_\mu - R_\mu)/2$, and comprise the $a_1(1260)$, $K_{1,A}$, $f_1(1285)$, and $f_1(1420)$. In the quark model, their quantum numbers are ${}^{2S+1}L_J = {}^{3}S_1$ and $\,^3P_1$, respectively. It is important to stress that the nature of axial-vector mesons is not yet clarified. As e.g.\ discussed in \cite{Schael:2005am,Wagner:2008gz,Leupold:2008is} the $a_1(1260)$ can be described as a dynamically generated state rather than a standard $\bar{q}q$ state. For other phenomenological studies of (axial-) vector mesons (with special emphasis on the mixing between the $K_{1,A}$ and the $K_{1,B}$ states) see \cite{Zhang:2014qza,Aaij:2014kwa,Liu:2014dxa,Hatanaka:2008gu,Ahmed:2011vr}.
	
	There are three direct mass terms for the homochiral vector multiplets,
		\begin{align}
			m_V^2\, \mathrm{tr} (L_\mu^2 + R_\mu^2) + \kappa\, \mathrm{tr} [ m^2_\mathrm{qk} (L_\mu^2 + R_\mu^2) ] \nonumber
			\\
			+ g_1^2 \mathrm{tr}[(\Phi^\dagger L^\mu - R^\mu \Phi^\dagger)(L^\mu \Phi - \Phi R^\mu)]\,. \label{mass_vector}
		\end{align}
	The first term, $\sim m_V^2$, is invariant under the full flavor group of $G_\mathrm{fl}$; note that parity requires the masses of the left and right handed fields to be equal. The second term involves a diagonal matrix proportional to the current quark masses, $m_\mathrm{qk} = \mathrm{diag}(m_\mathrm{u}\,, m_\mathrm{d}\,, m_\mathrm{s})$, and breaks the $SU(3)_\mathrm{V}$ symmetry for unequal quark masses. In diagonalizing the propagators it is necessary to include Eq.\ (\ref{vector_mixing}) as well; for the axial vector fields, this mixes the $a_1(1260)$ and derivatives of the pion, \textit{etc.}. The third term generates the (standard) mass difference between chiral partners. In fact, upon setting $\Phi\simeq\phi_{0},$ it reduces to $\phi^2_0 \mathrm{tr}(A_\mu^2)$, hence a mass term (proportional to the chiral condensate) is acquired by the axial-vector mesons only. Thus, the mass splitting among e.g.\ the $\rho(770)$ and the $a_1(1260)$ is generated by a non-vanishing condensate (just as for the splitting between the pion and $f_N$), see e.g.\ \cite{Ko:1994en,Urban:2001ru,Carter:1995zi} for a detailed discussion.
	
	Since these are homochiral fields, there is no term analogous to the $Z(3)_\mathrm{A}$ invariant terms for heterochiral fields, as in Eq.\ (\ref{anomalypseudoscalar}). Consequently, the mass eigenstates are naturally those of flavor, and not of $SU(3)_\mathrm{V}$: the $\rho(770)$ and $\omega(782)$ are dominantly composed of up and down quarks, and the $\phi(1020)$, mainly strange; the associated mixing angle between the pure strange and non-strange states is very small, $\theta_V \simeq -3.2^\circ$ \cite{pdg}. The same seems to hold for axial-vector mesons. Although the nature of these states is still subject to an ongoing debate, in particular concerning the amount of the meson-meson component in their wave functions \cite{Schael:2005am,Wagner:2008gz,Leupold:2008is,Roca:2005nm}, the states $f_{1}(1285)$ and $f_{1}(1420)$ appear to be predominantly non-strange and strange, respectively \cite{Divotgey:2013jba}. Beyond these complications which should be carefully addressed both theoretically and experimentally in the future, a small strange-nonstrange mixing of the underlying $\bar{q}q$ in the axial-vector channel is expected.\newline
		
	This is the basic observation which we now generalize to higher multiplets with spin one and two. Clearly, it is dependent upon the assumption that the chiral properties dominate even for these states. As can be anticipated, it generalizes immediately the results found for the above two examples: heterochiral states are mainly eigenstates of $SU(3)_\mathrm{V}$, and homochiral states, of flavor.\newline

\newpage

\onecolumngrid

\begin{center}
	\begin{table}[tbp]
	\caption{Chiral multiplets, their currents and transformations up to $J=3$. (${}^\star$ and/or $f_0 (1500)$. ${}^{\star\star}$a mix of.) The first two columns correspond to the assignment suggested in the `Quark Model' review of the PDG \cite{pdg}, to which we refer for further details and references (see also discussion in the text).}
	\label{tab:1}
	\centering
		\begin{tabular}{|c|p{3.6cm}|c|c|c|}
		\hline
		$J^{PC}\,,\, {}^{2S+1} L_J$ &
		$\begin{cases}
			I = 1\quad (\bar{u}d, \bar{d}u, \frac{\bar{d}d - \bar{u}u}{\sqrt{2}}) \\ 
			I = 1\quad (-\bar{u}s, \bar{s}u, \bar{d}s, \bar{s}d) \\ 
			I = 0\quad (\frac{\bar{u}u + \bar{d}d}{\sqrt{2}}, \bar{s}s)^{\star\star}
		\end{cases}$ &
		microscopic currents &
		chiral multiplet &
		\begin{minipage}{3cm}
			transformation under\\ $SU(3)_\mathrm{L} \times SU(3)_\mathrm{R} \times$ \\
			$\times U(1)_{\mathrm{A}}$
		\end{minipage} \\
		\hline\hline
		$0^{-+}\,,\, {}^1 S_0$ &
		$\begin{cases}
			\pi \\
			K \\
			\eta, \eta^\prime (958)
		\end{cases}$ &
		$P^{ij} = \frac{1}{2} \bar{q}^j \mathrm{i} \gamma^5 q^i$ &
		\multirow{5}{*}{\begin{minipage}{2.5cm}
			$\Phi = S + \mathrm{i}P$\\ $(\Phi^{ij} =\bar{q}^j_\mathrm{R} q^i_\mathrm{L})$
		\end{minipage}} &
		\multirow{5}{*}{$\Phi \longrightarrow \mathrm{e}^{-2\mathrm{i}\alpha} U_\mathrm{L} \Phi
		U_\mathrm{R}^\dagger$} \\
		$0^{++} \, , \, {}^3 P_0$ &
		$\begin{cases}
			a_0 (1450) \\
			K_0^\ast (1430) \\
			f_0 (1370), f_0 (1710)^\star
		\end{cases}$ &
		$S^{ij} = \frac{1}{2} \bar{q}^j q^i$ &
		& \\
		\hline
		$1^{--} \, , \, {}^1 S_1$ & $%
		\begin{cases}
		\rho(770) \\ 
		K^\ast (892) \\ 
		\omega(782), \phi(1020)%
		\end{cases}%
		$ & $V_\mu^{ij} = \frac{1}{2} \bar{q}^j \gamma_\mu q^i$ & %
		\begin{minipage}{2.5cm} $L_\mu = V_\mu + A_\mu$\\ $(L_\mu^{ij} =
		\bar{q}^j_\mathrm{L} \gamma_\mu q^i_\mathrm{L})$ \end{minipage} & $L_\mu
		\longrightarrow U_\mathrm{L} L_\mu U_\mathrm{L}^\dagger$ \\ 
		$1^{++} \, , \, {}^3 P_1$ & $%
		\begin{cases}
		a_1(1260) \\ 
		K_{1,A} \\ 
		f_1(1285), f_1(1420)%
		\end{cases}%
		$ & $A_\mu^{ij} = \frac{1}{2} \bar{q}^j \gamma^5 \gamma_\mu q^i$ & %
		\begin{minipage}{2.5cm} $R_\mu = V_\mu - A_\mu$\\ $(R_\mu^{ij} =
		\bar{q}^j_\mathrm{R} \gamma_\mu q^i_\mathrm{R})$ \end{minipage} & $R_\mu
		\longrightarrow U_\mathrm{R} R_\mu U_\mathrm{R}^\dagger$ \\ \hline
		$1^{+-} \, , \, {}^1 P_1$ & $%
		\begin{cases}
		b_1 (1235) \\ 
		K_{1,B} \\ 
		h_1 (1170), h_1 (1380)%
		\end{cases}%
		$ & $P_\mu^{ij} = - \frac{1}{2} \bar{q}^j\gamma^5 \overleftrightarrow{%
		D_\mu} q^i$ & \multirow{5 }{*}{\begin{minipage}{2.5cm} $\Phi_\mu =
		S_{\mu} + \mathrm{i} P_\mu$\\ $(\Phi_\mu^{ij} = \bar{q}^j_\mathrm{R}
		\mathrm{i} \overleftrightarrow{D_\mu} q^i_\mathrm{L})$ \end{minipage}}
		& \multirow{5}{*}{$\Phi_\mu \longrightarrow \mathrm{e}^{-2\mathrm{i}\alpha}
		U_\mathrm{L} \Phi_\mu U_\mathrm{R}^\dagger$} \\ 
		$1^{--} \, , \, {}^3 D_1$ & $%
		\begin{cases}
		\rho(1700) \\ 
		K^\ast (1680) \\ 
		\omega(1650), \phi(?)%
		\end{cases}%
		$ & $S_{\mu}^{ij} = \frac{1}{2} \bar{q}^j \mathrm{i} \overleftrightarrow{%
		D_\mu} q^i$ & & \\ \hline
		$2^{++} \, , \, {}^3 P_2$ & $%
		\begin{cases}
		a_2 (1320) \\ 
		K_2^\ast (1430) \\ 
		f_2 (1270), f_2^\prime (1525)%
		\end{cases}%
		$ & $V_{\mu\nu}^{ij} = \frac{1}{2} \bar{q}^j (\gamma_\mu \mathrm{i} 
		\overleftrightarrow{D_\nu} + \ldots) q^i$ & \begin{minipage}{4cm}
		$L_{\mu\nu} = V_{\mu\nu} + A_{\mu\nu}$\\ $(L_{\mu\nu}^{ij} =
		\bar{q}^j_\mathrm{L} (\gamma_\mu \mathrm{i}
		\overleftrightarrow{D_\nu} + \ldots) q^i_\mathrm{L})$ \end{minipage}
		& $L_{\mu\nu} \longrightarrow U_\mathrm{L} L_{\mu\nu} U_\mathrm{L}^\dagger$
		\\ 
		$2^{--} \, , \, {}^3 D_2$ & $%
		\begin{cases}
		\rho_2 (?) \\ 
		K_2 (1820) \\ 
		\omega_2 (?), \phi_2 (?)%
		\end{cases}%
		$ & $A_{\mu\nu}^{ij} = \frac{1}{2} \bar{q}^j (\gamma^5 \gamma_\mu \mathrm{i} 
		\overleftrightarrow{D_\nu} + \ldots) q^i$ & \begin{minipage}{4cm}
		$R_{\mu\nu} = V_{\mu\nu} - A_{\mu\nu}$\\ $(R_{\mu\nu}^{ij} =
		\bar{q}^j_\mathrm{R} (\gamma_\mu \mathrm{i}
		\overleftrightarrow{D_\nu} + \ldots) q^i_\mathrm{R})$ \end{minipage}
		& $R_{\mu\nu} \longrightarrow U_\mathrm{R} R_{\mu\nu} U_\mathrm{R}^\dagger$
		\\ \hline
		$2^{-+} \, , \, {}^1 D_2$ & $%
		\begin{cases}
		\pi_2 (1670) \\ 
		K_2 (1770) \\ 
		\eta_2 (1645), \eta_2 (1870)%
		\end{cases}%
		$ & $P_{\mu\nu}^{ij} = - \frac{1}{2}\bar{q}^j (\mathrm{i}\gamma^5 
		\overleftrightarrow{D_\mu} \overleftrightarrow{D_\nu} +
		\ldots) q^i$ & \multirow{5 }{*}{\begin{minipage}{4cm} $\Phi_{\mu\nu} =
		S_{\mu\nu} + \mathrm{i} P_{\mu\nu}$\\ $(\Phi_{\mu\nu}^{ij} =
		\bar{q}^j_\mathrm{R} (\overleftrightarrow{D_\mu}
		\overleftrightarrow{D_\nu} + \ldots) q^i_\mathrm{L})$ \end{minipage}}
		& \multirow{5}{*}{$\Phi_{\mu\nu} \longrightarrow
		\mathrm{e}^{-2\mathrm{i}\alpha} U_\mathrm{L} \Phi_{\mu\nu}
		U_\mathrm{R}^\dagger$} \\ 
		$2^{++} \, , \, {}^3 F_2$ & $%
		\begin{cases}
		a_2 (?) \\ 
		K_2^\ast (?) \\ 
		f_2 (?), f_2^\prime (?)%
		\end{cases}%
		$ & $S_{\mu\nu}^{ij} = -\frac{1}{2}\bar{q}^j (\overleftrightarrow{%
		D_\mu} \overleftrightarrow{D_\nu} + \ldots) q^i$ & &
		\\ \hline
		$3^{--} \, , \, {}^3 D_3$&
		$\begin{cases}
		\rho_3 (1690) \\
		K_3^\ast (1780) \\
		\omega_3 (1670), \phi_3 (1850)
		\end{cases}$ & $\vdots$ & $\vdots$ & $\vdots$ \\
		\hline
		\end{tabular}
	\end{table}
\end{center}

	\twocolumngrid

	(In order to make the following argumentation as clear as possible, we provide Tab.\ I, where we list all chiral multiplets, their quark-anti-quark currents, their transformation under $G_\mathrm{fl} \times U(1)_\mathrm{A}$, and their assignments to physical fields.)\newline
	
	
	\textit{Heterochiral vectors} -- Instead of inserting a Dirac matrix between a quark and anti-quark, we can use a gauge-covariant derivative $D_\mu = \partial_\mu - \mathrm{i} g G_\mu$,
		\begin{align}
			\Phi_\mu^{ij} \equiv \bar{q}_\mathrm{R}^{j} (\overleftrightarrow{D_\mu} + \ldots ) q_\mathrm{L}^i\,, \label{hetero_vector}
		\end{align}
	where $G_\mu$ is the $SU(3)_\mathrm{C}$ gauge vector field (the gluon). As it is necessary to pair a left-handed quark with a right-handed anti-quark, $\Phi_\mu$ is a heterochiral field, which transforms exactly like $\Phi$ in Eq.\ (\ref{chiral_transf_phi}),
		\begin{align}
			\Phi_\mu \longrightarrow \mathrm{e}^{-\mathrm{i}\alpha}\, U_\mathrm{L}\, \Phi_\mu\, U_\mathrm{R}^\dagger\,.
		\end{align} 
	Similarly, it is natural to form anomalous terms which are invariant under $G_\mathrm{fl}$ and $Z(3)_\mathrm{A}$ but not under $U(1)_\mathrm{A}$. Three terms analogous to Eq.\ (\ref{anomalypseudoscalar}) are
		\begin{align}
			\mathcal{L}_{\Phi_\mu}^{\text{anomaly}} = &- b_\mathrm{A}^{(1)} [ \mathrm{tr} ( \Phi \times \Phi_\mu \cdot  \Phi^\mu ) + \mathrm{c.c.} ]  \nonumber
			\\
			&- b_\mathrm{A}^{(2)} [ \mathrm{tr} ( \Phi \times \partial_\mu \Phi \cdot \Phi^\mu ) + \mathrm{c.c.}]  \nonumber
			\\
			&- b_\mathrm{A}^{(3)} [ \mathrm{tr} ( \Phi \times \Phi \cdot \Phi_\mu ) - \mathrm{c.c.} ]^2 + \ldots \label{pvlag}
		\end{align}
	plus other terms at sixth order, which we do not list here, and with the notation
		\begin{align}
			(A \times B)^{i i^\prime} = \frac{1}{3!} \epsilon^{i j k} \epsilon^{i^\prime j^\prime k^\prime} A^{j j^\prime} B^{k k^\prime}.
		\end{align}
	
	The vector multiplet decomposes into $\Phi_\mu = S_\mu + \mathrm{i} P_\mu$ \cite{Giacosa:2016hrm}. The $P^\mu$'s have $J^{PC} = 1^{+-}$, and may be the $b_1(1235)$, $K_{1,B}$, $h_1(1170)$, and $h_1(1380)$; in the quark model, they have ${}^{2S+1} L_J = {}^{1} P_1$. The $S^\mu$'s, with $J^{PC} = 1^{--}$, may be the $\rho(1700)$, $K^\ast(1680)$, $\omega(1650)$, and an excited $\phi(?)$ \cite{Piotrowska:2017rgt}; in the quark model, they have ${}^{2S+1} L_J = {}^{3} D_1$. Many of these states are not well understood: for example, the $h_1(1380)$ is still omitted from the standard assignment in the Particle Data Group \cite{pdg}.
	
	Our predictions for these states are clear: If the first term in Eq.\ (\ref{pvlag}) dominates, both heterochiral vectors should be close to eigenstates of $SU(3)_\mathrm{V}$, and not diagonal in flavor. The second term contributes to decay channels for one $\Phi_\mu$ to two $\Phi$'s. The third term only contributes to the $P_\mu$ states, $\sim - b_\mathrm{A}^{(3)} \phi_0^4 (h_{1,0}^{\mu})^{2}$ [$h_{1,0}^\mu$ is the singlet, a mixture of $h_1(1170)$ and $h_1(1380)$]. The complete form of the effective Lagrangian will allow one to make detailed predictions for the mixing angles driven by the anomaly. As for the heterochiral scalars and tensors (see Eq.\ (\ref{mixing_angle}) below) we expect a non-negligible negative mixing angle for the heterochiral vectors.
	
	
	\textit{Homochiral tensors} -- Multiplets with spin two are formed by combining Dirac matrices and covariant derivatives. If we use one of each, we obtain left and right handed fields,
		\begin{align}
			L_{\mu\nu}^{ij} & \equiv \bar{q}_\mathrm{L}^j (\gamma_\mu \overleftrightarrow{D_\nu} + \gamma_\nu \overleftrightarrow{D_\mu} + \ldots) q_\mathrm{L}^i\,, \nonumber
			\\
			R_{\mu\nu}^{ij} & \equiv \bar{q}_\mathrm{R}^j (\gamma_\mu \overleftrightarrow{D_\nu} + \gamma_\nu \overleftrightarrow{D_\mu} + \ldots) q_\mathrm{R}^i\,.
		\end{align}
	These are homochiral fields, which are invariant under $U(1)_\mathrm{A}$, and transform under $G_\mathrm{fl}$ as the homochiral vectors in Eq.\ (\ref{homo_vector_transf}),
		\begin{align}
			&L_{\mu\nu} \longrightarrow U_\mathrm{L}\, L_{\mu\nu}\, U_\mathrm{L}^\dagger\,, &&R_{\mu\nu} \longrightarrow U_\mathrm{R}\, R_{\mu\nu}\, U_\mathrm{R}^\dagger\,.
		\end{align}
	As for the homochiral vector multiplet, the only terms involving the anomaly involve at least two derivatives.
	
	The states $V_{\mu\nu} = (L_{\mu\nu} + R_{\mu\nu})/2$ with $J^{PC} = 2^{++}$ include the $a_2(1320)$, $K_{2}^\ast(1430)$, $f_2(1270)$, and $f_2^\prime(1525)$, with ${}^{2S+1} L_J = {}^{3} P_2$ in the quark model. There is also $A_{\mu \nu}$ with $J^{PC} = 2^{--}$, with ${}^{2S+1} L_J = {}^{3} D_2$ in the quark model. These states are not well measured, except for the $K_2(1820)$. They are presumably orbital excitations of the $a_1(1260)$, $K_{1,A}$, $f_1(1285)$, and $f_1(1420)$. For a general phenomenological overview of tensor mesons, see \cite{Chen:2011qu,Anisovich:2004vj,Wang:2014yza,Molina:2008jw,Geng:2008gx,Jido:2003cb}.
	
	Our analysis predicts that the effects of the anomaly are small, and that states are eigenstates of flavor. This agrees with experiment \cite{pdg} and recent lattice studies \cite{Briceno:2017qmb}, where the $f_{2}(1270)$ and $f_{2}^{\prime }(1525)$ correspond, to a very good approximation, to unmixed non-strange and strange states, with a small mixing angle $\theta_T \simeq 3.2^\circ$ \textcolor{blue}{\cite{pdg}}. In general, masses and decays fit very nicely into the quark-nonet paradigm without any effect of the anomaly \cite{Burakovsky:1997ci,Cirigliano:2003yq,Giacosa:2005bw}.
	
	
	\textit{Heterochiral tensors} -- We cannot use two Dirac matrices to form spin-two mesons, since the product of two Dirac matrices is $\gamma_\mu \gamma_\nu = g_{\mu\nu} - \mathrm{i} \sigma_{\mu\nu}$. The first term is a scalar, and equivalent to previous fields. As $\sigma_{\mu\nu}$ is antisymmetric, this reduces to a vector field (for details see \cite{Leupold:2005ep,Terschluesen:2010ik}).
	
	We can also use two covariant derivatives,
		\begin{align}
			\Phi_{\mu\nu}^{ij} \equiv \bar{q}_\mathrm{R}^j (\overleftrightarrow{D_\mu} \overleftrightarrow{D_\nu} + \overleftrightarrow{D_\nu} \overleftrightarrow{D_\mu} + \ldots) q_\mathrm{L}^i\,.
		\end{align}
	To have spin two, we must take the traceless part of $\Phi_{\mu\nu}$ \cite{Koenigstein:2016tjw}. This field is heterochiral, and transforms like $\Phi$, with
		\begin{align}
			\Phi_{\mu\nu} \longrightarrow \mathrm{e}^{-\mathrm{i}\alpha}\, U_\mathrm{L}\, \Phi_{\mu\nu}\, U_\mathrm{R}^\dagger \,.
		\end{align}
	Terms which are invariant under $G_\mathrm{fl} \times Z(3)_\mathrm{A}$, but not under $U(1)_\mathrm{A}$ begin with
		\begin{align}
			\mathcal{L}_{\Phi_{\mu\nu}}^{\text{anomaly}} = &- c_\mathrm{A}^{(1)} [ \mathrm{tr} ( \Phi \times \Phi_{\mu \nu} \cdot \Phi^{\mu \nu} ) + \mathrm{c.c.} ] \nonumber
			\\
			&- c_\mathrm{A}^{(2)} [ \mathrm{tr} ( \partial_\mu \Phi \times  \partial_\nu \Phi \cdot \Phi^{\mu\nu} ) + \mathrm{c.c.} ]  \nonumber
			\\
			&- c_\mathrm{A}^{(3)} [ \mathrm{tr} ( \Phi \times \Phi \cdot \Phi_{\mu\nu} ) - \mathrm{c.c.} ]^2 + \ldots \label{ptlag}
		\end{align}
	in direct analogy with Eqs.\ (\ref{anomalypseudoscalar}) and (\ref{pvlag}).
	
	The parity eigenstates are given by $\Phi_{\mu\nu} = S_{\mu\nu} + \mathrm{i} P_{\mu\nu}$. The odd parity states, the $P_{\mu\nu}$, have $J^{PC} = 2^{-+}$ and ${}^{2S+1} L_J = {}^{1} D_2$ in the quark model. Candidates are the $\pi_2(1670)$, $K_2(1770)$, $\eta_2(1645)$, and $\eta_2(1870)$ \cite{Wang:2014sea,Bing:2013fva,Anisovich:2010nh}. The even parity states, the $S_{\mu\nu}$, have $J^{PC}=2^{++}$ and ${}^{2S+1} L_J = {}^{3} F_2$ in the quark model. These are not well known experimentally, and should be some sort of $a_2$, $K_2^\ast$, $f_2$, and $f_2^\prime$ states (which are orbital excitations of the $\rho(1700)$, $K^\ast(1680)$, $\phi(?)$, and $\omega(1650)$).
	
	As before, the first term contributes to masses for $\Phi_{\mu\nu}$, and will lead to the multiplet being close to eigenstates of $SU(3)_\mathrm{V}$. The second term contributes to decays of one $\Phi_{\mu \nu}$ to two $\Phi$'s.
	
	The third term affects only pseudo-tensor mesons and delivers a contribution to the mass of the singlet state proportional to
		\begin{align}
			- \gamma_\mathrm{A} (\eta_{2,0,\mu\nu})^2 & = - \frac{\gamma_\mathrm{A}}{3} (\sqrt{2} \eta_{2,N,\mu\nu} + \eta_{2,S,\mu\nu})^2\,,
		\end{align}
	with $\gamma_\mathrm{A} \simeq c_\mathrm{A}^{(2)}\phi_0^4.$ Indeed a phenomenological study of known experimental widths and ratios \cite{Koenigstein:2016tjw} finds $\theta_{PT}\simeq -42^\circ$, hence a surprisingly large mixing,
		\begin{align}
			\begin{pmatrix}
			\eta_2(1645)\\
			\eta_2(1870)
			\end{pmatrix} = \begin{pmatrix}
			\cos \theta_{PT} & \sin \theta_{PT}\\
			-\sin \theta_{PT} & \cos \theta_{PT}
			\end{pmatrix} \begin{pmatrix}
			\eta_{2,N}\\
			\eta_{2,S}
			\end{pmatrix}\,,
		\end{align}
	where $\eta_{2,N} = (\bar{u}u + \bar{d}d)/\sqrt{2}$ and $\eta_{2,S} = \bar{s}s$. This result can be nicely explained by the presence of the axial anomaly in that sector. Moreover, considering only the third term in Eq.\ (\ref{ptlag}) the corresponding mixing angle $\theta_{PT}$ turns out to be negative (for realistic values of $\gamma_\mathrm{A}$) just as for heterochiral scalars,
		\begin{equation}
			\theta_{PT} \simeq -\frac{1}{2} \arctan \left[ \frac{2 \sqrt{2} \gamma_\mathrm{A}}{(m_{K_2(1770)}^2 - m_{\pi_2(1670)}^2 - \gamma_\mathrm{A})}\right] \label{mixing_angle}
		\end{equation}
	This result is approximate, since we neglect $c_\mathrm{A}^{(1)}$, and other mixing effects which are suppressed for a large number of colors. A more precise determination is possible with more experimental input.
	
	\textit{Homochiral mesons with $J=3$} -- It is even possible to extend our considerations to higher spin. As a further example, upon introducing the $J=3$ homochiral objects,
		\begin{align}
			L_{\mu\nu\rho}^{ij} & \equiv \bar{q}_\mathrm{L}^j (\gamma_\mu D_\nu D_\rho + \ldots) q_\mathrm{L}^i\,, \nonumber
			\\
			R_{\mu\nu\rho}^{ij} & \equiv \bar{q}_\mathrm{R}^{j}(\gamma_\mu D_\nu D_\rho + \ldots ) q_\mathrm{R}^i\,,
		\end{align}
	the nonet $V_{\mu\nu\rho} = ( L_{\mu\nu\rho} + R_{\mu\nu\rho} )/2$ corresponds to the rather well known mesons with $J^P =3^-$ (${}^{2S+1}L_J = {}^{3}D_{3}$) including the $\rho_{3}(1690)$, $K_3^\ast(1780)$, $\omega_3(1670)$, $\phi_3(1850)$. Similar to the pseudotensor mesons, the chiral partners of the $3^{--}$ are not yet known. The strange-nonstrange mixing angle leading to $\omega_3(1670)$ and $\phi_3(1850)$ is listed by the PDG as $\sim 3^\circ$ \cite{pdg}, hence very small, in agreement with our expectations for homochiral multiplets.
	
	
	\textit{Conclusions} -- Assuming that higher spin states can be categorized according to the unbroken chiral symmetry, they can be classified according to whether they are homo- or heterochiral states. The principal prediction is whether the flavor mixing of strangeness follows $SU(3)_\mathrm{V}$ symmetry, for heterochiral multiplets, or just flavor, for homochiral multiplets.
	
	Higher spin states are usually classified using models of constituent quarks. Such a framework will automatically yield multiplets classified according to flavor. Thus homochiral multiplets should agree with the predictions of quark models.
	
	In contrast, our analysis yields qualitatively new predictions for the heterochiral multiplets. There is evidence for this in the heterochiral tensor multiplet \cite{Koenigstein:2016tjw}.
	
	Besides masses and mixing, in the future one can study also various decay channels by using Eqs.\ (\ref{pvlag}) and (\ref{ptlag}). Similar studies can be performed with other approaches, such as lattice gauge theory \cite{Briceno:2017qmb}, chiral perturbation theory \cite{Molina:2008jw,Geng:2008gx,Jido:2003cb}, Schwinger-Dyson equations, and so on.
	Our predictions can be tested experimentally at the ongoing effort at BESIII \cite{Mezzadri:2015lrw,Marcello:2016gcn}, as well as GlueX \cite{Ghoul:2015ifw,Zihlmann:2010zz,Shepherd:2014hsa} and CLAS12 \cite{Rizzo:2016idq}, both of which will start measurements soon, and later with PANDA at FAIR \cite{Lutz:2009ff}.\newline
	
	
	The authors thank D.~H.~Rischke, Dennis~D.~Dietrich, the \textit{Chiral Field Theory Group}, E. Oset and C. Roberts for useful discussions. F.G. and R.D.P. thank the organizers of the 53rd Karpacz Winter School of Theoretical Physics, which engendered this collaboration. R.D.P. thanks E. Oset and C. Roberts for discussions at NSTAR 2017.  The work of F.~G.\ is supported by the Polish National Science Centre NCN through the OPUS project no.\ 2015/17/B/ST2/01625. R.D.P. thanks the U.S. Department of Energy for support under contract DE-SC0012704.
	
	
	\twocolumngrid

\end{document}